\documentstyle[epsf,aps]{revtex}

\begin{document}

\date{\today}

\title{Stochastic Wave-function Simulation of Two-time Correlation Functions}
\author{Timo Felbinger and Martin Wilkens}
\address{
  Institut f\"u{}r Physik, Universit\"a{}t Potsdam, 
  14469 Potsdam, Germany
  \newline
  http://www.quantum.physik.uni-potsdam.de
} 
\maketitle

\begin{abstract}
  We propose an optimized algorithm for the numerical simulation of two-time correlation
  functions by means of stochastic wave functions. 
  As a first application, we investigate the two-time correlation function of
  a nonlinear optical parametric oscillator.
\end{abstract}

\pacs{PACS-numbers: }

\section{Introduction}

 Spurred by the ever increasing speed of commercially available desk-top computers, 
 the analysis of dissipative quantum dynamics in terms of stochastic Schr\"o{}dinger 
 equation was recently promoted as an interesting alternative to the more traditional 
 approach which is based on the solution of the system's master equation. This 
 is because workstations are usually strong in in simulating a couple (say 100) 
 runs of a $N$-dimensional wave function, but --- because of memory limitations 
 --- perform rather weak in propagating the $O(N^{2})$ matrix elements of the 
 statistical operator \cite{molmer0}. In particular in quantum optics,
 where $N$ may be of the order $N=1000$ or larger, 
 stochastic techniques enjoy increasing popularity. 
 
 The technique is based on the ``unraveling'' \cite{carmichael} of a given 
 master equation 
 \begin{equation} \label{eq:master_def}
   \frac{d}{dt}\hat{\rho} \,=\, L\hat{\rho}
	 \,,
 \end{equation}
 in terms of a stochastic Schr\"o{}dinger equation, which
 propagates  a state vector $\psi_r(t)$ in such a way, 
 that the solution of the master equation (\ref{eq:master_def}), 
 $\hat\rho(t) = e^{Lt} \hat\rho(0)$, is recovered in the stochastic average,
 \begin{equation} \label{symmetricUnravel}
   \hat\rho(t) \,=\,
     \overline{|\psi_r(t)\rangle\langle\psi_r(t)|}
   \,.
 \end{equation}
 Here, the subscript $r$ labels a particular realization of the stochastic 
 process, and $\overline{(\ldots)}$ denotes the average over all realizations, 
 including a weighted sum over pure initial states. 

 A given master equation may be unraveled by a variety of stochastic methods, 
 which involve either appropriate generalizations of the continuous Wiener 
 stochastic processes, discontinuous jump processes, or a mixture of both. 
 The mathematical foundation of the various representations, their relation to 
 a specific physical set-up and their connection to the quantum measurement 
 problem is nicely reviewed in Refs.~\cite{zoller1,plenio}.
 For a semi-popular account, including some historical and philosophical issues,
 see Ref.~\cite{wilkens}.

 Even though the individual trajectories $|\psi_r(t)\rangle$ 
 enjoy great popularity for the illustration of a quantum systems' dynamics,
 the only physical quantities which may reliably be predicted using the set
 of $|\psi_r(t)\rangle$ are the time dependent expectation values of 
 quantum mechanical operators, 
 $\langle \hat{A}(t)\rangle \equiv {\rm tr} \left( \hat A \hat\rho(t) \right) $, viz.
 \begin{equation}
   \langle \hat A(t) \rangle 
   \,=\, 
   \overline{ \langle\psi_r(t)|\hat{A}|\psi_r(t)\rangle}
   \,.
 \end{equation}
 In particular, multi-time correlation functions may not be computed by means 
 of the representation (\ref{symmetricUnravel}).  
 The two-time correlation function 
 \begin{equation} \label{correlator}
   \langle \hat A(t)\hat B(0) \rangle 
   \,=\,
   {\rm tr} \left( \hat A e^{Lt}\hat B \hat\rho(0) \right)
   \,,
 \end{equation}
 for example, can not be evaluated by the recipe ``measure $\hat{B}$ in a 
 suitably selected initial state, propagate the post-measurement state 
 to time $t$, measure $\hat{A}$, repeat and average''. 
 The Heisenberg operators on the left hand side of 
 Eq.~(\ref{correlator}) must be treated with special care in open systems with
 non-unitary time evolution, as algebraic properties are generally not preserved
 by transformation to the Heisenberg picture 
 (i.\,e. $(\hat A \hat B)(t) \neq \hat A(t) \hat B(t)$)
 \cite{sondermann}.
 It turns out that for non-commuting $\hat A(t)$, $\hat B(0)$, the evaluation of
 Eq.~(\ref{correlator}) requires an
 intricate algorithm involving, at least, one pair of coupled stochastic 
 Schr\"o{}dinger equations. 

 Indeed, the correlator (\ref{correlator})
 may be viewed as the single-time expectation value of $\hat{A}$ with respect to 
 an improper state $\hat\chi(t)$, 
 $\left< \hat A(t) \hat B(0) \right> = {\rm tr}\left( \hat A \hat\chi(t) \right)$,
 where
 \begin{equation} \label{chiFormalSolution}
   \hat\chi(t) \,=\, e^{Lt} \hat B \hat\rho(0)
   \,.
 \end{equation}
 For a pure initial state, $\hat\rho(0) = |\psi(0)\rangle\langle\psi(0)|$,
 the initial value of $\hat\chi(t)$ is given in terms of a dyadic product of 
 two state vectors $|\psi(0)\rangle$ and 
 $|\phi(0)\rangle \equiv \hat B |\psi(0)\rangle$,
 \begin{equation}
   \hat\chi(0) \,=\, |\phi(0)\rangle\langle\psi(0)|
   \,.
 \end{equation}
 This indicates that the evaluation of (\ref{correlator}) may be reduced to the
 simulation of a single-time expectation value
 of $\hat A$ with respect to $\hat\chi(t)$, which comes, however,
 at the expense of dealing with two different wave functions simultaneously. 

 In the past, several stochastic wave function algorithms have been proposed 
 which support the numerical simulation of a two-time correlation function
 \cite{zoller1,molmer1,gisin3,breuer2}.
 Since all these algorithms are build to yield the correct result in the limit 
 of infinitely many runs, they are effective, but they are usually not very 
 efficient. The scheme proposed by Castin et al \cite{molmer1}, 
 for example, is threatened by 
 numerical instability as it relies on the subtraction of possibly large 
 numbers. The particular scheme proposed by Gardiner and Zoller \cite{zoller1}, 
 and Gisin \cite{gisin3}, on
 the other hand, is exponentially inefficient, that is the number of 
 trajectories which are needed for a reliable estimate of the desired 
 correlation function is bound from below by an {\em increasing} exponential 
 function of the correlator's time $t$. The somewhat more intricate algorithm 
 which may be extracted from 
 the work of Breuer and collaborators \cite{breuer2} 
 does not 
 suffer from this particular kind of ineffeciency, but it is not yet 
 optimized. To date no systematic investigation has addressed
 the issue of how to tailor an algorithm which is both effective and efficient.

 In the present paper we construct an optimal algorithm for a 
 class of stochastic representations which are characterized by pseudo-linear 
 Ito differential equations involving jump processes. The paper is organized 
 as follows. In Sec.~\ref{sectionSymmetric} we review the unraveling of the 
 master equation for a genuine statistical operator using Ito stochastic 
 calculus. In Sec.~\ref{sectionNonSymmetric}, we extend our analysis for the 
 stochastic representation of the dynamics of skew-symmetric state operators, 
 and we indicate an optimal algorithm.
 Using the simple example of spontaneous emission of a two-state system we 
 analyse the efficiency of alternative algorithm, including the Gardiner-Zoller 
 \cite{zoller1} and Breuer-Kappler-Petruccione \cite{breuer2} method. 
 As a non-trivial example, we apply our algorithm to the problem of tunneling in 
 the degenerate optical parametric oscillator.
 The methods of Castin and collaborators \cite{molmer1}, and of Breuer 
 and collaborators \cite{breuer2} are revied in Appendices~\ref{appendixBKP}
 and \ref{appendixMCD}.

\section{Propagation of proper state operators}
 \label{sectionSymmetric}

 In this section we seek the unraveling of the model master equation,
 \begin{equation} \label{simpleMasterEquation}
     \frac{d}{dt}\hat{\rho}
   \,=\,
     2 \hat \sigma \hat{\rho} \hat \sigma^\dagger
     - \hat \sigma^\dagger \hat \sigma \hat{\rho}
     - \hat{\rho}\hat \sigma^\dagger \hat \sigma
   \,,
 \end{equation}
 which -- dependent on the algebraic properties of the operators 
 $\hat{\sigma}$, $\hat{\sigma}^{\dagger}$ -- describes
 spontaneous emission of a 2-level atom or the damping of a cavity mode. 
 Generalizations to other systems will be covered in 
 Sec.~\ref{sectionGeneralization}.

 We unravel the master equation (\ref{simpleMasterEquation}) in terms of 
 an Ito stochastic differential equation
 \begin{equation} 
   \label{symmetricAnsatz}
    |d\psi_r(t)\rangle
    \,=\,
      d\xi(t) |\psi_r(t)\rangle 
      - d\mu(t) \hat{\sigma}^{\dagger}\hat{\sigma} | \psi_r(t)\rangle 
      + d\eta(t) \hat{\sigma} | \psi_r(t)\rangle
    \,,
 \end{equation}
 where $\xi(t)$, $\mu(t)$ and $\eta(t)$ are, in general complex, stochastic 
 processes, and differentials are forward oriented, 
 $d\psi(t) = \psi(t+dt)-\psi(t)$ etc. 
 Note that the equation is pseudo-linear, as the stochastic increments 
 $d\xi(t)$, $d\mu(t)$ and $d\eta(t)$ may depend on $\psi_r(t)$.

 A stochastic process like $\xi(t)$ is, in general, not differentiable and the 
 stochastic increment $d\xi(t)$ does not have the properties of an ordinary 
 differential, i.\,e.\ it can not, in general, be treated as ``infinitesimal''. 
 However, from Eq.~(\ref{symmetricUnravel}) and the assumed smoothness of 
 $\hat{\rho}(t)$ it follows that the stochastic average of the dyadic product 
 $|\psi_r(t)\rangle\langle\psi_r(t)|$ must be well-behaved and differentiable,
 \begin{equation}
   \label{eq:dSymmetricDyad}
   d \left( \overline{
      \left| \psi_r(t) \right\rangle \left\langle \psi_r(t) \right|
   } \right)
   \,=\,
   \overline{
      \left| d\psi_r(t) \right\rangle \left\langle \psi_r(t) \right|
   }
   + \overline{
      \left| \psi_r(t) \right\rangle \left\langle d\psi_r(t) \right|
   }
   + \overline{
      \left| d\psi_r(t) \right\rangle \left\langle d\psi_r(t) \right|
   }
   \,.
 \end{equation}
 Note that we are not allowed to drop the last term as would be the case if 
 $|d\psi_r(t)\rangle$ were an ordinary differential. 

 By virtue of Eq.~(\ref{symmetricUnravel}), the differential
 (\ref{eq:dSymmetricDyad}) equals the forward differential of the 
 density operator, $d\rho(t)=\rho(t+dt)-\rho(t)$. 
 For our particular model (\ref{simpleMasterEquation})
 \begin{equation}
   \label{eq:dRho}
   d\hat\rho(t)
   \,=\,
   2 \hat{\sigma}\hat{\rho}(t)\hat{\sigma}^\dagger\,dt
   - \hat{\sigma}^{\dagger}\hat{\sigma}\hat{\rho}(t)\,dt
   - \hat{\rho}(t)\hat{\sigma}^{\dagger}\hat{\sigma}\,dt\,.
 \end{equation}
 Since Eqs. (\ref{eq:dSymmetricDyad}) and (\ref{eq:dRho}) must coincide 
 regardless of the value of $\rho(t)$, they 
 must coincide for any individual member of the ensemble $\rho(t)$, say 
 $\psi_r(t)$. Hence we set $\rho(t)=|\psi_{r}(t)\rangle\langle\psi_{r}(t)|$ 
 in Eq.~(\ref{eq:dRho}), and in Eq.~(\ref{eq:dSymmetricDyad}) consider 
 $\psi_{r}(t)$ as fixed. 
 As we make no assumptions on the action of the operators $\hat{\sigma}$ and 
 $\hat{\sigma}^{\dagger}$, respectively, we insert the stochastic ansatz 
 (\ref{symmetricAnsatz}) 
 into Eq.~(\ref{eq:dSymmetricDyad}) and compare coefficients of the right hand 
 sides of Eqs. (\ref{eq:dSymmetricDyad}) and (\ref{eq:dRho}). 
 The result reads
 \begin{eqnarray}
   \label{symmetricMoments}
  &&  \overline{ \left(
        d\xi(t) + d\xi^\ast(t) + d\xi(t) d\xi^\ast(t) 
      \right) }  
        \, = \, 0
  \cr &&
      \overline{ \left(
        ( d\xi(t) + 1 ) d\mu^\ast(t) 
      \right) }
        \, = \, dt 
    \,,\quad
      \overline{ \left(
        ( d\xi(t) + 1 ) d\eta^\ast(t) 
      \right) }^{\vphantom{M^M}}
        \, = \, 0  
  \cr &&
      \overline{ \left(
        d\mu(t) d\mu^\ast(t) 
      \right) }
        \, = \, 0  
    \,,\quad
      \overline{ \left(
        d\mu(t) d\eta^\ast(t) 
      \right) }
        \, = \, 0
    \,,\quad
      \overline{ \left(
        d\eta(t) d\eta^\ast(t) 
      \right) }^{\vphantom{M^M}}
        \, = \, 2dt  
    \,,
  \end{eqnarray}
 where $\overline{\vphantom{\psi}\ldots}$ denotes an average over the 
 stochastic increments for fixed $\psi_r(t)$. 

 The unraveling (\ref{symmetricUnravel})
 conserves the norm square $\vert\psi_r(t)\vert^{2}$ in stochastic average, as 
 is obvious when taking the trace in Eq.~(\ref{symmetricUnravel}),
 \begin{equation}
   \overline{
      \left\langle \psi_r(t) | \psi_r(t) \right\rangle 
   }
   \,=\, 1
   \,.
 \end{equation}
 In fact, all well-established algorithms preserve the norm even for every 
 single realization\footnote{
   Sometimes this is not obvious at first glance: in some recipies for 
   unraveling (\ref{simpleMasterEquation}), norm conservation may be 
   violated temporarily; 
   however, explicit renormalization will then become necessary, 
   at the latest when expectation values are computed.
 }:
 \begin{equation}
   \label{strictNormConservation}
    \forall\, t,\, r : \quad
    \left\langle \psi_r(t) | \psi_r(t) \right\rangle 
    \,=\, 1
    \,.
 \end{equation}
 As a rule, algorithms which have been derived from some theory of continuous 
 measurement respect norm conservation in the strong 
 version~(\ref{strictNormConservation}), as they deal with proper
 states which admit a physical interpretation at any time.

 If numerical efficiency rather than merely effectiveness is an issue, 
 algorithms which respect the stronger 
 conditon~(\ref{strictNormConservation}) ought to be preferred. This is 
 because they generate trajectories of equal weigth, avoiding waste of
 CPU time on simulating a large number of trajectories when only a few
 ``heavy'' ones actually contribute to the result. A more formal proof
 of this requirement is given in Sec.~\ref{ourDerivation}.

 The strong conservation law~(\ref{strictNormConservation}) imposes additional
 constraints on the stochastic increments. Taking the stochastic differential
 of Eq. (\ref{strictNormConservation}) we obtain
 \begin{equation}
     \langle\psi_r(t)|d\psi_r(t)\rangle
     + \langle d\psi_r(t)|\psi_(t)\rangle
     + \langle d\psi_r(t)|d\psi_(t)\rangle
   \,=\,
     0
   \,,
 \end{equation}
 which --- using the stochastic ansatz (\ref{symmetricAnsatz}) and dropping the
 time arguments for readability --- amounts to the following constraint:
 \begin{eqnarray}
   \label{strictNormConstraints}
   d\xi + d\xi^\ast - (d\mu + d\mu^\ast)Q + d\eta P + d\eta^\ast P^\ast
     + d\xi^\ast d\xi - d\xi^\ast d\mu Q + d\xi^\ast d\eta P
 \cr 
     - d\mu^\ast d\xi Q + d\mu^\ast d\mu S - d\mu^\ast d\eta R
     + d\eta^\ast d\xi P^\ast - d\eta^\ast d\mu R^\ast + d\eta^\ast d\eta Q
   &\, =\, & 0\,,
 \end{eqnarray}
 where $P := \langle \psi_r | \hat\sigma | \psi_r \rangle$, 
       $Q := \langle \psi_r | \hat\sigma^\dagger \hat\sigma | \psi_r \rangle$, 
       $R := \langle \psi_r | \hat\sigma^\dagger \hat\sigma \hat\sigma| \psi_r \rangle$, 
 and   $S := \langle \psi_r | \hat\sigma^\dagger \hat\sigma 
                              \hat\sigma^\dagger \hat\sigma| \psi_r \rangle$.

 Even if supplemented by the strong condition of norm 
 conservation~(\ref{strictNormConservation}), the 
 conditions~(\ref{symmetricMoments}) and (\ref{strictNormConstraints}) still offer 
 quite some freedom in the choice of a correspondig stochastic model. 

 Most prominent are the models of quantum state diffusion \cite{gisin0}, where 
 the stochastic Schr\"o{}dinger equation assumes the form of a Langevin 
 equation with Wiener stochastic increments, and models of quantum jumps, 
 where the time evolution is mostly deterministic, only interrupted by 
 discontinuous jumps at random times. Since jump methods seem to be favoured in 
 numerical applications we concentrate on the jump processes in the following.

 In a simple two-branch jump process, at any time $t$, the tripel 
 $\left( d\xi, d\mu, d\eta \right)$ can take on one of two different values,
 \begin{eqnarray}
   \label{generalJumpAnsatz}
     (d\xi, d\mu, d\eta) \,=\, \left\lbrace
       \matrix{  (\xi_{\rm J}, \mu_{\rm J}, \eta_{\rm J}) 
                   & ,\quad \hbox{with probability }\, dp \hfill
                \cr (d\xi_{\rm c}, d\mu_{\rm c}, 0 )  
                   & ,\quad \hbox{with probability }\, 1 - dp \hfill   }
     \right.
 \end{eqnarray}
 Most of the time, $d\eta$ is zero while $d\xi$ and $d\mu$ take on 
 well-defined, infinitesimal, values, so that the state evolves continuously,
 denoted by the index~${\rm c}$. 
 With a certain probability $dp \propto dt$, a ``jump'' 
 occurs, corresponding to a finite-valued tripel 
 $(\xi_{\rm J}, \mu_{\rm J}, \eta_{\rm J})$.
 Such processes are often derived from a theory of measurement, where
 jumps correspond to ``clicks'' of some photo detector.

 Note that while Eq.~(\ref{symmetricAnsatz}) contains stochastic differentials
 $d\xi$, $d\mu$, $d\eta$, we are now left with 6 parameters
 $\xi_{\rm J}$, $\mu_{\rm J}$, $\eta_{\rm J}$,
 $d\xi_{\rm c}$, $d\mu_{\rm c}$, and $dp$, 
 which are deterministic functions of the state vector 
 $\left| \psi_r(t) \right\rangle$;
 the only stochastic element in (\ref{generalJumpAnsatz}) is the decision which 
 branch to take. 
 We have retained the differential notation for the parameters 
 $d\xi_{\rm c}$ and $d\mu_{\rm c}$, which will 
 be of order $dt$, in contrast to 
 $\xi_{\rm J}$, $\mu_{\rm J}$ and $\eta_{\rm J}$, which will be of order 1.

 The most simple jump process is obtained by choosing $\xi_{\rm J} = -1$, 
 $\mu_{\rm J} = 0$; from this, and the 
 strong condition of norm conservation (\ref{strictNormConservation}), one 
 readily derives the remaining parameters
 \begin{equation}
     \eta_{\rm J} \,=\,
        \left| \hat\sigma \left| \psi_r \right\rangle \right|^{-1} 
      \,,\quad
     dp \,=\,  2 dt \left| \hat\sigma | \psi_r \rangle \right|^2
       \,,\quad
     d\xi_{\rm c} \,=\, {dp\over 2}
       \,,\quad
     d\mu_{\rm c} \,=\, dt
     \,.
 \end{equation}
 The stochastic process (\ref{symmetricAnsatz}) now reads
 \begin{eqnarray} \label{simpleSymmetricJump}
   \left| d\psi_r \right\rangle &\,=\,& \left\lbrace
       \matrix{
         \left(
           - \hbox{\rm 1\hskip-0.25em l} 
           + \left| \hat\sigma \left| \psi_r \right\rangle \right|^{-1} 
               \hat \sigma 
         \right) \left| \psi_r \right\rangle
          ,&\, dp 
       \cr
         \left(
           \left\langle \psi_r | \hat\sigma^\dagger\hat\sigma | \psi_r \right\rangle
               - \hat \sigma^\dagger\hat\sigma 
         \right) dt \, \left| \psi_r \right\rangle 
         ,&\,  1 - dp 
         \,,
       }
     \right.
   \end{eqnarray}
 and the vector $|\psi_r(t)\rangle$ will be mapped according to
 $|\psi_r(t)\rangle \mapsto |\psi_r(t+dt)\rangle$,
 \begin{eqnarray} 
   \lineskip=0pt
   |\psi_r(t+dt)\rangle \,=\, \left\lbrace  \matrix{ 
        \noalign{ \hrule width 1pt height 0pt }
     \cr
         \left| \hat \sigma \left| \psi_r(t) \right\rangle \right|^{-1} 
           \hat\sigma \left| \psi_r(t) \right\rangle
        ,&\,
           dp = 2 dt \left\langle \psi_r(t) | \hat\sigma^\dagger \hat\sigma | \psi_r(t) \right\rangle
     \cr \noalign{ \hrule width 0pt height 3pt }
     \cr
       {\textstyle
          \left( \hbox{\rm 1\hskip-0.25em l} -  dt \hat\sigma^\dagger \hat\sigma \right) 
        \over \textstyle
           \left| 
              \left( \hbox{\rm 1\hskip-0.25em l} - dt \hat\sigma^\dagger \hat\sigma \right) 
              \left| \psi_r(t) \right\rangle 
           \right|
       } 
       \left| \psi_r(t) \right\rangle
       ,&\,  1 - dp
     \cr \noalign{ \hrule width 1pt height 0pt }
     \cr
   } \right.
 \end{eqnarray}
 This algorithm and its generalizations are quite popular in the quantum 
 optics community, where they have been
 applied to a variety of problems, see Refs.~[1-6].

\section{Propagation of non-symmetric operators}
  \label{sectionNonSymmetric}

 \subsection{Construction of an efficient general-purpose algorithm}
   \label{ourDerivation}
   \label{subsectionSkewProblem}
 
  As indicated in the introduction, numerical simulation of the two-time 
  correlation function $\langle A(t)B(0)\rangle$ requires the unraveling of the 
  skew-symmetric operator 
    $\hat{\chi}(t) \,=\, e^{Lt}\hat{B}\hat{\rho}(0)$.
  Here, a naive generalization of recipes like (\ref{simpleSymmetricJump}) 
  can not succeed,
  as there is no obvious way to define the jump probability for an operator
  like $\hat\chi(t)$ which is neither definite nor hermitian.
 
  Recall, however, that the unraveling is only required for a pure initial 
  state, $\hat\rho(0)=|\psi(0)\rangle\langle\psi(0)|$, since the case of mixed 
  initial state is easily obtained by means of a suitable weighted sum over such pure 
  states. Denoting $|\phi(0)\rangle=\hat{B}|\psi(0)\rangle$, one has  
  $\hat\chi(0)=|\phi(0)\rangle\langle\psi(0)|$, which is the dyadic product of two 
  state vectors $\phi$ and $\psi$. Thus unraveling may well proceed along the 
  lines of Sec.~\ref{sectionSymmetric}, this time however for a pair 
  $   \left( \left| \phi_r(t) \right\rangle
    , \left| \psi_r(t) \right\rangle \right) $ 
  of vectors, such that the skew-symmetric $\hat\chi(t)$ is recovered in the
  stochastic average:
  \begin{eqnarray}
    \label{skewUnraveling}
    \hat\chi(t)
    &\,=\,&
    \overline{|\phi_r(t)\rangle\langle\psi_r(t)|}
    \,,
  \cr
    \left\langle A(t) B(0) \right\rangle
    &\,=\,&
    \overline{ \left\langle \psi_r(t) | \hat A | \phi_r(t) \right\rangle }
    \,.
  \end{eqnarray}
 

  \smallskip

  We start the unraveling of $\hat\chi(t)$ from an ansatz similar 
  to (\ref{symmetricAnsatz}), but now for a pair of wave functions:
  \begin{eqnarray} \label{skewAnsatz}
     \left| d\phi_r \right\rangle &=&
       d\xi_1 \left| \phi_r \right\rangle 
       - d\mu_1 \hat\sigma^\dagger \hat\sigma \left| \phi_r \right\rangle 
       + d\eta_1 \hat\sigma \left| \phi_r \right\rangle 
       \,,
   \cr
     \left| d\psi_r \right\rangle &=&
       d\xi_2 \left| \psi_r \right\rangle 
       - d\mu_2 \hat\sigma^\dagger \hat\sigma \left| \psi_r \right\rangle 
       + d\eta_2 \hat\sigma \left| \psi_r \right\rangle 
       \,.
  \end{eqnarray}
  Analogous to Eq.~(\ref{symmetricMoments}), we can again derive a set of 
  necessary conditions for the stochastic increments:
  \begin{mathletters}
  \begin{eqnarray} \label{skewMoments}
       \hfill
         \overline{ 
           \left( d\xi_1 + d\xi_2^\ast + d\xi_1 d\xi_2^\ast
         \right) }
       &=& 0 \,, 
       \label{skewMomentsXiXi}
     \\
       \hfill
         \overline{ ( d\xi_1 + 1 ) d\mu_2^\ast } \,\,=
           \overline{ ( d\xi_2^\ast + 1 ) d\mu_1 }     
         &=& dt \,,
        \label{skewMomentsXiMu}
       \strut^{\vphantom{M^M}}
     \\
       \hfill
       \overline{ ( d\xi_1 + 1 ) d\eta_2^\ast } \,=\,
         \overline{ ( d\xi_2^\ast + 1 ) d\eta_1 } 
        &=& 0 \,,
        \label{skewMomentsXiEta}
       \strut^{\vphantom{M^M}}
     \\
       \hfill
         \overline{ d\mu_1 d\mu_2^\ast } &=& 0 \,,
          \label{skewMomentsMuMu}
       \strut^{\vphantom{M^M}}
     \\
       \hfill
       \overline{ d\mu_1 d\eta_2^\ast } \,=\,
         \overline{ d\mu_2^\ast d\eta_1 } &=& 0 \,,
        \label{skewMomentsMuEta}
       \strut^{\vphantom{M^M}}
     \\
       \hfill
       \overline{ d\eta_1 d\eta_2^\ast } &=& 2 dt 
        \label{skewMomentsEtaEta}
       \strut^{\vphantom{M^M}}
  \end{eqnarray}
  \end{mathletters}
  We restrict ourselves to a jump process similar to (\ref{simpleSymmetricJump}),
  but with free parameters:
  \begin{eqnarray}
    \lineskip=0pt
    \baselineskip=0pt
    \label{ourJumpAnsatz}
       \left( \matrix{   \left| d\phi_r \right\rangle
                     \cr \left| d\psi_r \right\rangle } \right)
     &=& 
       \left\lbrace \matrix{
         \noalign{ \hrule width 0pt height 10pt }
       \cr
           \left( \matrix{
             \left( - \hbox{\rm 1\hskip-0.25em l} 
                    + \eta_{1,\rm J} \hat\sigma  \right)
             \left| \phi_r \right\rangle
           \cr
             \left( - \hbox{\rm 1\hskip-0.25em l} 
                    + \eta_{2,\rm J} \hat\sigma  \right)
             \left| \psi_r \right\rangle  
           } \right)
         \,,&\quad dp
       \cr
         \noalign{ \hrule width 0pt height 4pt }
       \cr
         \left( \matrix{
           \left( d\xi_{1,\rm c}
                  - d\mu_{1,\rm c} \hat\sigma^\dagger\hat\sigma \right)
           \left| \phi_r \right\rangle
         \cr
           \left( d\xi_{2,\rm c}
                  - d\mu_{2,\rm c} \hat\sigma^\dagger\hat\sigma \right)
           \left| \psi_r \right\rangle
         } \right)
         \,,&\quad 1 - dp
       \cr 
         \noalign{ \hrule width 1pt height 0pt }
       \cr
       }\right. 
  \end{eqnarray}
  As above, the indices ${\rm c}$ and ${\rm J}$ denote the 
  continuous branch and the jump branch, respectively. 
  The ansatz (\ref{ourJumpAnsatz}), 
  although less general than in Eq. (\ref{skewAnsatz}), still contains enough degrees 
  of freedom to allow for a wide range of jump algorithms,  including many of those
  suggested in the literature \cite{zoller1,molmer1,gisin3}; it has the advantage of 
  beeing easy to implement, and it will automatically fullfill 
  the conditions (\ref{skewMomentsXiEta}) and (\ref{skewMomentsMuEta}).

  \smallskip
  There are two degrees of freedom in the choice of the free parameters
  in (\ref{ourJumpAnsatz}) which are mere gauge freedoms with no relevance to 
  the efficiency of the process (this has already been observed by Di\'osi, 
  see \cite{gisin3}):
  for some complex number $c \neq 0$, the transformation
  \begin{eqnarray} \label{skewGauge1}
    ( d\xi_{1,\rm c}, d\mu_{1,\rm c}, d\xi_{2,\rm c}, d\mu_{2,\rm c} ) 
    \mapstochar\longrightarrow
       ( c^\ast d\xi_{1,\rm c}, c^\ast d\mu_{1,\rm c}
        , c^{-1} d\xi_{2,\rm c}, c^{-1} d\mu_{2,\rm c} ) 
  \end{eqnarray}
  will leave (\ref{skewMoments}) invariant and leads to a process which is 
  equivalent in the sense that it predicts identical trajectories for all
  observable quantities.
  Similarly, there is another gauge freedom for 
  $\eta_{1,\rm J}$, $\eta_{2,\rm J}$: for any complex number $\tilde c \neq 0$,
  the transformation
  \begin{eqnarray} \label{skewGauge2}
    ( \eta_{1, \rm J}, \eta_{2,\rm J} ) \mapstochar\longrightarrow
     ( \tilde c^\ast \eta_{1,\rm J}, \tilde c^{-1} \eta_{2,\rm J} )
  \end{eqnarray}
  leads to an equivalent stochastic process.

  We are therefore free to choose the symmetric gauge:
  \begin{eqnarray} \label{ourGauge}
    \forall \,r, t\colon\quad
        \left| \phi_r(t) \right|
      \mathop=\limits^{!}
        \left| \psi_r(t) \right|
    \,.
  \end{eqnarray}
  As this must hold for both the continuous and the jump branch in 
  Eq.~(\ref{ourJumpAnsatz}), it fixes the modulus of both gauge parameters 
  $c$ and $\tilde c$ in (\ref{skewGauge1}) and (\ref{skewGauge2}). 
  The phase can be fixed by demanding that $d\xi_{1,2,\rm c}$ and 
  $\eta_{1,2,\rm J}$ be real numbers;
  from (\ref{skewMomentsXiMu}) it follows that $d\mu_{1,2,\rm c}$, too, 
  will be then real-valued.

  We are now left with 7 real-valued parameters to be determined; 
  since (\ref{skewMomentsXiEta}) and (\ref{skewMomentsMuEta}) are
  always satisfied by our ansatz (\ref{ourJumpAnsatz}), we have to fullfill 
  the following 7 conditions:
  \begin{eqnarray}
    \matrix{
      \hfill
      (1-dp)(d\xi_{1,\rm c} d\xi_{2,\rm c} 
                  + d\xi_{1,\rm c} + d\xi_{2,\rm c} ) & = & dp \hfill
        & (\ref{skewMomentsXiXi})
    \cr \hfill
      (1-dp)( d\xi_{1,\rm c} + 1 ) d\mu_{2,\rm c} & = & dt \hfill
        & (\ref{skewMomentsXiMu})
      \strut^{\vphantom{M^M}}
    \cr \hfill
      (1-dp)(d\xi_{2,\rm c} + 1) d\mu_{1,\rm c} & = & dt \hfill
        & (\ref{skewMomentsXiMu})
      \strut^{\vphantom{M^M}}
    \cr \hfill
      (1-dp)( d\mu_{1,\rm c} d\mu_{2,\rm c} ) & = & 0 \hfill
        & (\ref{skewMomentsMuMu})
      \strut^{\vphantom{M^M}}
    \cr \hfill
      \eta_{1,\rm J} \eta_{2,\rm J} dp & = & 2 dt \hfill
        & (\ref{skewMomentsEtaEta})
      \strut^{\vphantom{M^M}}
    \cr \hfill
      \left| \eta_{1,\rm J} \hat\sigma \left| \phi_r \right\rangle \right| 
       & = & 
      \left| \eta_{2,\rm J} \hat\sigma \left| \psi_r \right\rangle \right| \hfill
      & (\ref{ourGauge})
      \strut^{\vphantom{M^M}}
    \cr \hfill
        \left|
          \left( ( d\xi_{1,\rm c} + 1 ) - d\mu_{1,\rm c} \hat\sigma^\dagger\hat\sigma ) \right)
          \left| \phi_r \right\rangle
        \right| 
      & = & 
        \left|
          \left( ( d\xi_{2,\rm c} + 1 ) - d\mu_{2,\rm c} \hat\sigma^\dagger\hat\sigma ) \right)
          \left| \psi_r \right\rangle
        \right| 
      \hfill
      & (\ref{ourGauge})
      \strut^{\vphantom{M^M}}
    }
  \end{eqnarray}
  Taking into account that $d\xi_{1,\rm c}$, $d\xi_{2,\rm c}$, 
  $d\mu_{1,\rm c}$, $d\mu_{2,\rm c}$ and $dp$ are quantities of
  order $dt$, while $\eta_{1,\rm J}$ and $\eta_{2,\rm J}$ will be quantities 
  of order 1, these conditions can be simplified: the values
  \begin{equation}
    d\mu_{1,\rm c} =  d\mu_{2,\rm c} = dt
  \end{equation}
  are already determined, 
  and we are left with just 4 constraints
  for the 5 remaining parameters $dp$, $d\xi_{1,\rm c}$, $d\xi_{2,\rm c}$,
  $\eta_{1,\rm J}$, $\eta_{2,\rm J}$:
  \begin{eqnarray}  \label{remainingConstraints}
    \matrix{
      \hfill
        d\xi_{1,\rm c} + d\xi_{2,\rm c} & = & dp \hfill
          & (\ref{skewMomentsXiXi})
      \strut^{\vphantom{M^M}}
    \cr \hfill
      \eta_{1,\rm J} \eta_{2,\rm J} dp & = & 2 dt \hfill
        & (\ref{skewMomentsEtaEta})
      \strut^{\vphantom{M^M}}
    \cr \hfill
      \left| \eta_{1,\rm J} \hat\sigma \left| \phi_r \right\rangle \right| 
       & = & 
      \left| \eta_{2,\rm J} \hat\sigma \left| \psi_r \right\rangle \right| \hfill
      & (\ref{ourGauge})
      \strut^{\vphantom{M^M}}
    \cr \hfill
      \left|
        \left( ( d\xi_{1,\rm c} + 1 ) - d\mu_{1,\rm c} \hat\sigma^\dagger\hat\sigma ) \right)
        \left| \phi_r \right\rangle
      \right| & = & \left|
        \left( ( d\xi_{2,\rm c} + 1 ) - d\mu_{2,\rm c} \hat\sigma^\dagger\hat\sigma ) \right)
        \left| \psi_r \right\rangle
      \right| 
      \hfill
      & (\ref{ourGauge})
      \strut^{\vphantom{M^M}}
    }
  \end{eqnarray}
  We are thus free to impose exactly one additional arbitrary constraint; 
  since all conditions listed above are either required (within our ansatz for 
  the jump process) or mere gauge conditions with no effect on the trajectories
  of observable quantities,
  it will be this single condition which will determine the efficiency of the 
  algorithm.


 \subsection{Minimizing the error of an jump algorithm}

  In a simulation of $K$ trajectories 
    $ \hat\chi_j := \left| \phi_j \right\rangle \left\langle \psi_j \right| $,
    $ j = 1,\ldots,K $,
  the true skew operator $\hat\chi$ will be approximated by
  \begin{equation}
    \tilde\chi \,=\, {1\over K} \sum_{j=1}^K \hat\chi_j
    \,.
  \end{equation}
  For any correct algorithm, we have
  \begin{equation}
    \overline{ \tilde\chi }  \,=\, \hat\chi
    \,.
  \end{equation}
  At the same time, we would like the error to be minimal:
  \begin{equation} \label{desiredCondition0}
    \overline{ \left\Vert \tilde\chi - \chi \right\Vert^2 }
      \,\mathop=\limits^{!} \, {\rm minimal}
      \,.
  \end{equation}
  If we use the norm 
    $ \left\Vert \chi \right\Vert^2 := {\rm tr} \left( \chi^\dagger \chi \right) $,
  then we find for our ensemble of $K$ trajectories:
  \begin{eqnarray} \label{error1}
    \overline{ \left\Vert \tilde\chi - \chi \right\Vert^2 }
    &=&
      {1\over K^2} \sum_{j,\,l=1}^K
        \overline{ {\rm tr}\left( \hat\chi_j^{\dagger} \hat\chi_l \right) }
      - {\rm tr} \left( \hat\chi^\dagger \hat\chi \right)
      \,.
  \end{eqnarray}
  For independend trajectories $\hat\chi_j$ and $\hat\chi_l$, where $j\neq l$,
  we have
  \begin{eqnarray}
    \overline{ {\rm tr}\left( \hat\chi_j^\dagger \hat\chi_k \right) }
    & = &
      \sum_{r,\,r'} P(r) P(r') 
        {\rm tr}\left( \hat\chi_r^\dagger \hat\chi_{r'} \right) 
  \cr
    & = &
        {\rm tr}\left( 
          \left( \sum_r P(r) \hat\chi_r^\dagger \right)
          \left( \sum_{r'} P(r') \hat\chi_{r'} \right) 
        \right)
    \,=\, 
        {\rm tr} \left( \hat\chi^\dagger \hat\chi \right)
    \,,
  \end{eqnarray}
  where we sum over all realizations $r$ of the stochastic process and
  denote with $P(r)$ the probability of the realization $r$.
  Thus, Eq.~(\ref{error1}) becomes:
  \begin{eqnarray} \label{errorEstimate}
    \overline{ \left\Vert \tilde\chi - \chi \right\Vert^2 }
    &=&
      {1\over K^2} \left(
        \sum_{j=1}^K
          \overline{ {\rm tr}\left( \hat\chi_j^{\dagger} \hat\chi_j \right) }
      + \sum_{j,l=1 \atop j\neq k}^K
          \overline{ {\rm tr}\left( \hat\chi_j^{\dagger} \hat\chi_l \right) }
      \right)
      - {\rm tr} \left( \hat\chi^\dagger \hat\chi \right)
  \cr &=&
      {1\over K} \overline{ {\rm tr}\left( \hat\chi_r^{\dagger} \hat\chi_r \right) }
      + {K-1 \over K} {\rm tr} \left( \hat\chi^\dagger \hat\chi \right)
      - {\rm tr} \left( \hat\chi^\dagger \hat\chi \right)
  \cr &=&
      {1^{\vphantom{M}} \over K} 
         \overline{ {\rm tr}\left( \hat\chi_r^{\dagger} \hat\chi_r \right) }
      - {1 \over K} {\rm tr} \left( \hat\chi^\dagger \hat\chi \right)
    \,.
  \end{eqnarray}
  The condition (\ref{desiredCondition0}) now reads:
  \begin{equation} \label{desiredCondition1}
     \overline{ {\rm tr}\left( \hat\chi_r^\dagger \hat\chi_r \right) }
     \,\equiv\,
     \overline{
       \left\langle \phi_r | \phi_r \right\rangle
       \left\langle \psi_r | \psi_r \right\rangle
     }
      \, \mathop=\limits^! \, {\rm minimal}
      \,.
  \end{equation}
  By minimizing
  \begin{equation} 
    \sum_r P(r) {\rm tr} \left( \hat\chi_r^\dagger \hat\chi_r \right)
  \end{equation} 
  under the constraints
  \begin{eqnarray}
    \sum_r P(r) \,=\, 1
    \qquad \hbox{and} \qquad
    \sum_r P(r) \hat\chi_r \,=\, \hat\chi
    \,,
  \end{eqnarray}
  (\ref{desiredCondition1}) turns out to be equivalent to
  \begin{equation} \label{desiredCondition2}
    \forall\,r,r':\qquad
       {\rm tr} \left( \hat\chi_r^\dagger \hat\chi_r \right) 
      \, \mathop=\limits^! \,
       {\rm tr} \left( \hat\chi_{r'}^\dagger \hat\chi_{r'} \right)
    \,.
  \end{equation}
  We see that it is indeed desirable to avoid algorithms containing rare 
  but ``large'' trajectories. 
  Unfortunately, ensuring Eqs.~(\ref{desiredCondition1}) or (\ref{desiredCondition2})
  will be in general impossible, so we replace it by a weaker condition which is 
  necessary but not sufficient for Eq. (\ref{desiredCondition1}) to hold, but which will
  nevertheless suffice as the last missing constraint required by the one remaining
  degree of freedom in (\ref{remainingConstraints}):
  \begin{equation} \label{ourCondition}
    \forall \, \left| \phi \right\rangle, \left| \psi \right\rangle: \quad
      {d\over dt} \overline{ \left(
        \left\langle \phi | \phi \right\rangle
        \left\langle \psi | \psi \right\rangle
      \right)
    } \, \mathop=\limits^{!} \: {\rm minimal}
    \,.
  \end{equation}
  Here, the ensemble average $\overline{\phantom{MM}}$ is taken only
  over the two possible branches in (\ref{ourJumpAnsatz}), not over an ensemble 
  of possible states before performing the time step: we want this condition to
  be fullfilled for every member of the ensemble individually. 
  We will show below that this will even guarantee a monotonical decrease of the
  norm of both $|\phi\rangle$ and $|\psi\rangle$ in every single realization.

  We can now proceed to derive a jump algorithm fullfilling (\ref{ourCondition}).
  The time derivative can be evaluated separately for the jump branch and
  the continuous branch in (\ref{ourJumpAnsatz}); for the jump branch we find:
  \begin{eqnarray}
    d \left. \left( \left\langle \phi | \phi \right\rangle
                   \left\langle \psi | \psi \right\rangle \right) \right|_{\rm J}
    & = &
    \eta_{1,\rm J}^2 \eta_{2,\rm J}^2
       \left\langle \phi | \hat\sigma^\dagger\hat\sigma | \phi \right\rangle
       \left\langle \psi | \hat\sigma^\dagger\hat\sigma | \psi \right\rangle
     -   \left\langle \phi | \phi \right\rangle
         \left\langle \psi | \psi \right\rangle 
      \strut^{\vphantom{M^M}}
  \cr & = &
    { 4 dt^2 \over dp^2 }^{\vphantom{M}} \Phi^2 \Psi^2 - s^2
     \,,
  \end{eqnarray}
  where
  \begin{eqnarray}
	  \label{algAbbr}
      s  \,:= \, \left\langle \phi \vert \phi \right\rangle 
               \,\equiv\, \left\langle \psi \vert \psi \right\rangle
      \:, \quad
      \Phi \, := \, 
        \left\vert \hat\sigma \left\vert \phi \right \rangle \right\vert
      \:, \quad
      \Psi \,:=\,  
        \left\vert \hat\sigma \left\vert \psi \right \rangle \right\vert
      \,.
  \end{eqnarray}

  For the continuous branch, we have:
  \begin{eqnarray}
    d \left. \left( \left\langle \phi | \phi \right\rangle
                   \left\langle \psi | \psi \right\rangle \right) \right|_{\rm c}
     & = &
        2\left( d\xi_{1,\rm c} + d\xi_{2.\rm c} \right) 
                   \left\langle \phi | \phi \right\rangle
                   \left\langle \psi | \psi \right\rangle 
      \strut^{\vphantom{M^M}}
  \cr && \qquad
      - 2 dt \left(
          \left\langle \phi | \hat\sigma^\dagger\hat\sigma | \phi \right\rangle
          \left\langle \phi | \phi \right\rangle
        + 
          \left\langle \psi | \hat\sigma^\dagger\hat\sigma | \psi \right\rangle
          \left\langle \psi | \psi \right\rangle 
      \right)
      \strut^{\vphantom{M^M}}
  \cr & = &
        2\left( d\xi_{1,\rm c} + d\xi_{2.\rm c} \right) s^2
        - 2dt \left( \Phi^2 + \Psi^2 \right) s
      \strut^{\vphantom{M^M}}
  \cr & = &
        2 dp s^2 - 2dt \left( \Phi^2 + \Psi^2 \right) s
      \:.
      \strut^{\vphantom{M^M}}
  \end{eqnarray}

  So we have on average:
  \begin{eqnarray}
      \left\langle
        d \left(
          \left\langle \phi | \phi \right\rangle
          \left\langle \psi | \psi \right\rangle
        \right)
      \right\rangle_r 
    & = &
      dp \, d \left. \left( \left\langle \phi | \phi \right\rangle
                \left\langle \psi | \psi \right\rangle \right) \right|_{\rm J}
      + (1 - dp)
            d \left. \left( \left\langle \phi | \phi \right\rangle
                   \left\langle \psi | \psi \right\rangle \right) \right|_{\rm c}
  \cr & = &
    { 4 dt^2 \over dp } \Phi^2 \Psi^2 + dp \, s^2
        - 2dt \left( \Phi^2 + \Psi^2 \right) s
    \:.
  \end{eqnarray}
  Minimizing this with respect to the only remaining parameter $dp$ yields
  our choice for the jump probability $dp$:
  \begin{equation} \label{ourJumpProbability}
      dp \,=\, \displaystyle { 2 \over s } \Phi \Psi dt 
      \:.
  \end{equation}

  The remaining parameters can now readily be derived from (\ref{remainingConstraints}):
  \begin{eqnarray}
    \label{ourParameters}
    \matrix{ 
      \eta_{1,\rm J} \!&=&\! \displaystyle { \sqrt{s} \over \Phi }
      \,, \hfill & \:
      \eta_{2,\rm J} \!&=&\! \displaystyle { \sqrt{s} \over \Psi } 
      \,, \hfill
      \strut^{\vphantom{M^M}}
    \cr \strut^{\vphantom{M^M}}
      d\xi_{1,\rm c} \!&=&\! \displaystyle { 1 \over s }
        \left( \Phi \Psi + {1\over 2} \Phi^2 - {1\over 2} \Psi^2 \right) dt
      \,, \hfill & \:
      d\xi_{2,\rm c} \!&=&\! \displaystyle { 1 \over s }
        \left( \Phi \Psi
           - {1\over 2} \Phi^2 + {1\over 2} \Psi^2 \right) dt
      \,,
    \cr \strut^{\vphantom{M^M}}
      d\mu_{1,\rm c} \!&=&\! dt
      \,, \hfill & \:
      d\mu_{2,\rm c} \!&=&\! dt
      \,. \hfill
    }
  \end{eqnarray}

 \subsection{Comparison with other algorithms}

  We are going to compare several alternative algorithms by  analyzing
  the radiative decay of a two-level atom.
  The atom is prepared in its excited state $\left| e \right\rangle$ 
  from where it decays into its ground state $\left| g \right\rangle$ by 
  means of spontaneous emission. The corresponding master equation is given 
  by Eq.~(\ref{simpleMasterEquation}) with the specification 
  $\hat\sigma=\sqrt{\gamma}|g\rangle\langle e|$. 

  We wish to compute the 
  correlation function
  \begin{eqnarray}
     g(t) &=& \left\langle \hat\sigma^\dagger(t) \hat\sigma \right\rangle
     \,,
  \end{eqnarray}
  which for this particular problem should result in $g(t)=e^{-\gamma t}$.

  Starting with the initial state
  \begin{eqnarray}
     \left( \left| \phi(0) \right\rangle , \left| \psi(0) \right\rangle \right)
       &=&  
     \left( \left| g \right\rangle , \left| e \right\rangle \right)
     \,,
  \end{eqnarray}
  we observe that the continuous branch in Eq.~(\ref{ourJumpAnsatz}) will leave
  the state unchanged except for scalar factors changing the norm of $\phi_r$
  and $\psi_r$.
  The first jump maps $|\phi_r(t)\rangle$ onto $0$. As this value can never change 
  again, only trajectories with no jump up to time $t$ will contribute to $g(t)$. 

  \subsubsection{The Gardiner-Zoller algorithm}

   In Ref.~\cite{zoller1} an algorithm is suggested which is characterized by the 
   following choice of parameters:
   \begin{eqnarray}
     \label{GZparameters}
     dp &=& 2 \Psi^2 dt 
     \,,\quad
     \eta_{1,\rm J} \,=\, 
     \eta_{2,\rm J} \,=\, \displaystyle{ 1\over\Psi }
     \,,
   \cr \strut^{\vphantom{M^M}}
     d\xi_{1,\rm c} &=& 
     d\xi_{2,\rm c} \,=\, \Psi^2 dt
     \,,\quad 
     d\mu_{1,\rm c } \,=\, d\mu_{2,\rm c} \,=\, dt 
     \,,
   \end{eqnarray}
	 with $\Psi = \left| \hat\sigma|\psi\rangle \right|$, see Eq.~(\ref{algAbbr}).
   Note that this algorithm displays a certain asymmetry in favour of
   $\left| \psi_r \right\rangle$, as it will keep 
   $\left| \psi_r \right\rangle$ strictly normalized, but exerts little control 
   over the norm of $\left| \phi_r \right\rangle$. This may 
   pose a serious problem, as the following analysis demonstrates.
 
   Using the parameters (\ref{GZparameters}), trajectories with no jump evolve 
   according to
   \begin{eqnarray}
     \label{gardinerExponential}
      \left( \left| \phi_{\rm nJ}(t) \right\rangle 
             , \left| \psi_{\rm nJ}(t) \right\rangle \right)
        &=&  
      \left( \left| {\rm g} \right\rangle  e^{\gamma t}
             , \left| {\rm e} \right\rangle \right)
      \,,
   \end{eqnarray}
   and the probability of getting a trajectory without jumps up to time $t$ is
   \begin{eqnarray}
      p_{\rm nJ} &=& e^{ -2\gamma t}
      \,.
   \end{eqnarray}
   The expectation value
   \begin{eqnarray}
      g(t) &=& 
        p_{\rm nJ} 
          \left\langle \psi_{\rm nJ}(t) | \hat\sigma^\dagger 
                 | \phi_{\rm nJ}(t) \right\rangle
       \,=\, e^{-\gamma t}
   \end{eqnarray}
   is predicted correctly, but it requires the simulation of at least
   \begin{eqnarray} \label{NGZ}
      N_{\rm GZ} &\gg& p_{\rm nJ}^{-1} \,=\, e^{2\gamma t}
   \end{eqnarray}
   trajectories to get a statistically significant result.
   In view of the exponential growth of the right-hand side of Eq.~(\ref{NGZ}),
	 the algorithm (\ref{GZparameters}) is not suited for investigation of the
	 long-time dynamics of $g(t)$.

  \subsubsection{The doubled-Hilbertspace method}
 
   A different choice of parameters which is symmetric with respect to 
   $\left| \psi_r \right\rangle$ and $\left| \phi_r \right\rangle$ 
   is suggested in Ref.~\cite{breuer2}:
   \begin{eqnarray}
     \label{BreuerParameters}
     && dp \,=\, 2 dt \left(
        \left\langle \psi_r | \hat\sigma^\dagger \hat\sigma | \psi_r \right\rangle 
      + \left\langle \phi_r | \hat\sigma^\dagger \hat\sigma | \phi_r \right\rangle \right)
     \,,\;
     \eta_{1,\rm J} \,=\, 
     \eta_{2,\rm J} \,=\, \left(
        \left\langle \psi_r | \hat\sigma^\dagger \hat\sigma | \psi_r \right\rangle 
      + \left\langle \phi_r | \hat\sigma^\dagger \hat\sigma | \phi_r \right\rangle 
      \right)^{-1}
     ,
   \cr \strut^{\vphantom{M^M}}
     && d\xi_{1,\rm c} \,=\, d\xi_{2,\rm c} 
     \,=\,   \left\langle \psi_r | \hat\sigma^\dagger \hat\sigma | \psi_r \right\rangle 
           + \left\langle \phi_r | \hat\sigma^\dagger \hat\sigma | \phi_r \right\rangle 
     \,,\; 
     d\mu_{1,\rm c } \,=\, d\mu_{2,\rm c} \,=\, dt 
     \,.
   \end{eqnarray}
   Note that in this formulation the normalization 
   $   \left\langle \psi_r | \psi_r \right\rangle 
     + \left\langle \phi_r | \phi_r \right\rangle \equiv 1 $, which is required
	 for $t=0$, will be preserved by the dynamics.
   A more detailed account of this algorithm is given in Appendix \ref{appendixBKP}.

   \smallskip

   For the example of the radiative decay of a 2-level atom, we find that
   in this algorithm, trajectories with no jumps evolve like
   \begin{eqnarray} \label{BreuerEvolution}
        \left( \matrix{
             \left| \phi_{\rm nJ}(t) \right\rangle 
         \cr \left| \psi_{\rm nJ}(t) \right\rangle 
        } \right)
      &=&  
        { 1 \over \sqrt{ 1 + e^{-2\gamma t}} }
        \left( \matrix{
              \left| g \right\rangle 
          \cr \left| e \right\rangle e^{-\gamma t}
        } \right)
      \,.
   \end{eqnarray}
   The probability to simulate such a trajectory is given by
   \begin{eqnarray}
      p_{\rm nJ} &=& {1\over 2} \left( 1 + e^{-2\gamma t} \right)
      \,.
   \end{eqnarray}
   Therefore, one needs
   \begin{eqnarray} \label{NBKP}
      N_{\rm BKP} &\gg& {2 \over 1 + e^{-2\gamma t} }
   \end{eqnarray}
   trajectories to predict $g(t)$. 
   
   The analysis of the algorithm proposed in \cite{molmer1}, which we review in 
   Appendix~\ref{appendixMCD},
   leads to an identical estimate for this example.
   This is no conincidence, as both algorithms are equivalent for any system where 
   $\hat\sigma|\phi_{\rm r} \rangle$ and $\hat\sigma |\psi_{\rm r}\rangle$ are always
   orthogonal.
   
   The estimate (\ref{NBKP}), although already much better than Eq.~(\ref{NGZ}), can still be 
   improved, as we shall now demonstrate.
 
  \subsubsection{The optimized algorithm}
   
   If we apply our algorithm, defined by the choice of parameters 
   (\ref{ourJumpProbability}) and (\ref{ourParameters}), to the example of the 
   spontaneously emitting 2-level atom, it turns out that the algorithm degenerates 
   and the evolution becomes deterministic: Trajectories with no jump evolve like
   \begin{eqnarray} \label{optimizedTrajectory}
     \left| \phi(t) \right\rangle \,=\, e^{-{\gamma\over 2}t} \left| g \right\rangle
   \,,\quad
     \left| \psi(t) \right\rangle \,=\, e^{-{\gamma\over 2}t} \left| e \right\rangle
   \,.
   \end{eqnarray}
   As $dp$ vanishes for all times, no jumps will ever occur. 
	 The trajectory (\ref{optimizedTrajectory})
   is the only possible one,
   and the value of the correlation function $g(t)$,
   \begin{eqnarray}
      g(t) &=& \left\langle \Psi(t) | \hat\sigma^\dagger | \Phi(t) \right\rangle
       \,=\, e^{-\gamma t}
      \,,
   \end{eqnarray}
   is predicted exactly by just this single trajectory.

 \subsection{Introduction of a ``no-jump-probability''}

  As it is desirable not to have to check for jumps in every elementary time
  interval $dt$, one needs a more efficient way to find the time of the next jump.

  We label the jump times by $\tau_1$, $\tau_2$, $\tau_3$, \ldots,
  and define the quantities $q_k(t),\, t > \tau_k$, by
  \begin{equation}
    q_k(\tau_k+t) \, := \,
      P\left( \mbox{ ``no jump in $]\tau_k, \tau_k+t]$'' } \right)
		\,,
  \end{equation}
  where $P$ denotes a probability.
  Clearly, the $q_k(t)$ are monotonically decreasing, starting
  from $q_k(\tau_k) = 1$, and they obey the differential equation
  \begin{eqnarray} \label{qDiffEquation}
    q_k(t+dt) & = & q_k(t) \left( 1 - dp(t) \right) 
    \:.
  \end{eqnarray}

  For the time $\tau_{k+1}$ of the next jump after the jump at $\tau_k$, we have
  for $t > \tau_k$,
  \begin{eqnarray}
    dP\left( \tau_{k+1} \in\, ] t, t+dt ] \right) 
      & = & q_k(t) dp(t) 
      \, = \, | dq_k(t) |
      \,,
  \end{eqnarray}
  (Probability of the jump not occuring before $t$, multiplied by the probability 
  of {\sl any} jump occuring in $dt$).
  If $r$ is a random number with uniform distribution in $[0, 1[$, then
  we find that for any $t>\tau_k$,
  \begin{eqnarray}
    dP\left( r \in [q_k(t), q_k(t)+dq_k(t) [ \right)
     \,=\, |dq_k(t)| 
     \,=\, dP\left( \tau_{k+1} \in ] t, t+dt ] \right) 
     \,.
  \end{eqnarray}
  The inverse function $q_k^{-1}(r)$ can therefore be used to map uniformely distributed
  random numbers $r \in [0,1[$ onto jump times $\tau_{k+1}$ with the correct
  distribution.
  This allows to simulate the process (\ref{ourJumpAnsatz}) by integrating
  equation (\ref{qDiffEquation}) in parallel to the continuous branch of
  (\ref{ourJumpAnsatz}), and performing a jump whenever $q_k(t)$ drops below a
  previously chosen random number $r \in [0,1[$.

 \subsection{Generalization for arbitrary master equations}
   \label{sectionGeneralization}

   If we generalize this to a master equation containing several loss 
   channels, as well as an Hamiltonian interaction, 
   \begin{eqnarray}
     {d\over dt} \hat\chi(t) &=&
       \sum_{k=1}^n \left(
         2 \hat\sigma_k \hat\chi(t) \hat\sigma_k^\dagger
           - \hat\sigma_k^\dagger \hat\sigma_k \hat\chi(t) 
           - \hat\chi(t) \hat\sigma_k^\dagger \hat\sigma_k
       \right)
       - i \left[ \hat H, \hat\chi(t) \right]
     \,,
   \end{eqnarray}
   with
   \begin{equation}
     \hat\chi(0) \,=\, \left| \phi(0) \right\rangle \left\langle \psi(0) \right|
     \,,
   \end{equation}
   we arrive at the following recipe for unraveling:

   \smallskip
   Initialization:

   \begin{itemize}
     \item 
		   Use a gauge transformation
             $\left| \phi(0) \right\rangle \mapsto c\left| \phi(0) \right\rangle$,
             $\left| \psi(0) \right\rangle \mapsto c^{-1}\left| \psi(0) \right\rangle$),
       such that $\left\langle \phi(0) | \phi(0) \right\rangle 
                \,=\, \left\langle \psi(0) \vert \psi(0) \right\rangle$;
     \item initialize the real-valued auxiliary variable $q(0) \,:=\, 1$;
     \item choose a random number $r \in [0,1[$.
   \end{itemize}

   The continuous evolution between jumps is governed by the following 
   pseudo-linear equations of motion:
   \begin{eqnarray}  
     \label{geometricEquationOfMotion}
     {d\over dt} q & = & - { 2 \over s }\, q \sum_{k=1}^n \Phi_k \Psi_k 
     \,,
   \cr {d\over dt} \left\vert \phi_r \right\rangle & = &
     {1 \over s} \sum_{k=1}^n
       \left( \Phi_k \Psi_k  + {1\over 2}\Phi_k^2 - {1\over 2}\Psi_k^2 \right)
       \left\vert \phi_r \right\rangle
       - \left( \sum_{k=1}^n \hat\sigma_k^\dagger \hat\sigma_k + i \hat H \right)
           \left\vert \phi_r \right\rangle
     \,,
   \cr {d\over dt} \left\vert \psi_r \right\rangle & = &
     {1 \over s} \sum_{k=1}^n
       \left( \Phi_k\Psi_k - {1\over 2}\Phi_k^2 + {1\over 2}\Psi_k^2 \right)
       \left\vert \psi_r \right\rangle
       - \left( \sum_{k=1}^n \hat\sigma_k^\dagger \hat\sigma_k + i \hat H \right)
           \left\vert \psi_r \right\rangle
     \,,
   \end{eqnarray}
   where, as above,
   \begin{eqnarray} \label{abbreviations}
       s  \,:= \, \left\langle \phi_r \vert \phi_r \right\rangle 
                \,\equiv\, \left\langle \psi_r \vert \psi_r \right\rangle
       \,, \quad
       \Phi_k \, := \, \left\vert \hat\sigma_k \left\vert \phi_r \right \rangle \right\vert
       \,, \quad
       \Psi_k \, :=\,  \left\vert \hat\sigma_k \left\vert \psi_r \right \rangle \right\vert
       \,.
   \end{eqnarray}
   Whenever the variable $q$ drops below the previously chosen random number $r$,
   a jump occurs according to the following scheme:
   \begin{itemize} 
     \item Choose randomly one jump operator $\hat\sigma_k$ using the 
           statistical weights $\Phi_k \Psi_k $ for the individual 
           channels;
     \item apply the following map:
       \begin{eqnarray} \label{jumpMapping}
              q & \mapstochar\longrightarrow & 1
       \,,
         \quad \left\vert \phi_r \right\rangle   \, \mapstochar\longrightarrow \,  
                 { \sqrt{s} \over \Phi_k } \hat\sigma_k \left\vert \phi_r \right\rangle 
       \,,
         \quad \left\vert \psi_r \right\rangle   \, \mapstochar\longrightarrow \,  
                 { \sqrt{s} \over \Psi_k } \hat\sigma_k \left\vert \psi_r \right\rangle 
       \,;
       \end{eqnarray}
     \item choose a new random number $r \in [0,1[$\,.
   \end{itemize}

  \subsection{Some properties of the proposed algorithm}

   From the map (\ref{jumpMapping}) it can be seen that jumps have no influence on the norms
   of $\left| \phi_r(t) \right\rangle$ and $\left| \psi_r(t) \right\rangle$
   (which is the local equivalent to the global condition (\ref{desiredCondition2})).
   The continuous evolution, however, will cause the norms to decrease montonically:
   \begin{eqnarray}
     {d\over dt} \left\langle \phi_r \vert \phi_r \right\rangle
     & = &  2 \left\langle \phi_r \right| \sum_{k=1}^n 
        \left(
          {1\over s} (\Phi_k \Psi_k + {1\over 2}\Phi_k^2 - {1\over 2}\Psi_k^2 )
            - \hat\sigma_k^\dagger \hat\sigma_k
        \right) \left| \phi_r \right\rangle
   \cr & = &
           2 \sum_{k=1}^n \left( \matrix{
                \noalign{ \hrule width 0pt height 7pt }
             \cr
               \sqrt{
                 \left\langle \phi_r | \hat\sigma_k^\dagger \hat\sigma_k | \phi_r \right\rangle
                 \left\langle \psi_r | \hat\sigma_k^\dagger \hat\sigma_k | \psi_r \right\rangle
               } \qquad\qquad
             \cr \quad - {1\over 2}
               \left( 
                   \left\langle \phi_r | \hat\sigma_k^\dagger \hat\sigma_k | \phi_r \right\rangle
                 +
                   \left\langle \psi_r | \hat\sigma_k^\dagger \hat\sigma_k | \psi_r \right\rangle
               \right)
             \cr
               \noalign{\hrule width 0pt height 0pt depth 1pt}
             \cr
           } \right)^{\vphantom{{M^M}}}
      \,.
   \end{eqnarray}
   From the properties of geometric and arithmetic means, we find:
   \begin{eqnarray} \label{monotonousDecay}
     {d\over dt} \left\langle \phi_r \vert \phi_r \right\rangle
     \,\equiv\, {d\over dt} \left\langle \psi_r \vert \psi_r \right\rangle
     \,\leq\, 0
     \,.
   \end{eqnarray}
   According to Eq. (\ref{errorEstimate}), the absolute error is bounded from above by
	 $K^{-1} \overline{ \langle\phi_r|\phi_r\rangle \langle\psi_r|\psi_r\rangle }$,
   and from Eq. (\ref{monotonousDecay}) it follows that this upper bound is
	 decreasing with time.
  
\begin{figure}
    \epsfxsize=120mm\epsffile{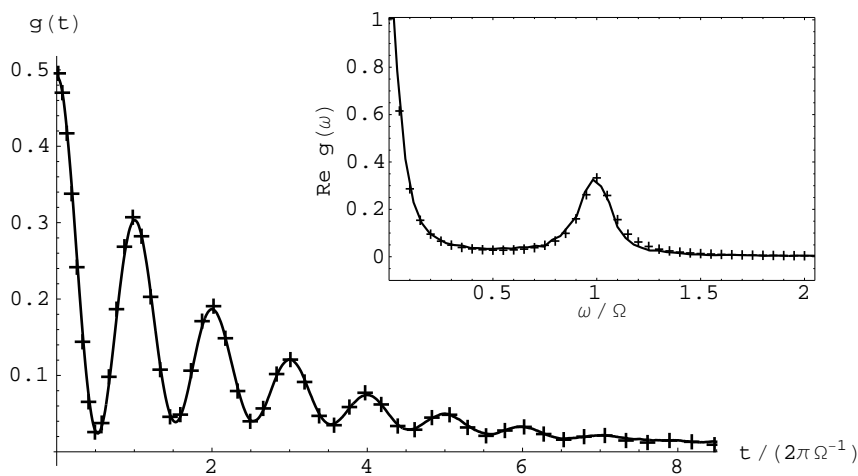}
    \caption{
      Time evolution of the correlation function 
			$g(t) = \langle \hat\sigma^\dagger(t) \hat\sigma \rangle$ for the 
      driven 2-level atom, with $\Omega = 8 \gamma$. 
			The inset shows the corresponding spectrum.
      The {\sl solid} line is the result of a Monte Carlo simulation
			with 5000 trajectories,
      while the {\sl crosses} represent exact analytical values.
    }
    \label{mollowgt}
  \end{figure}

   An implementation of our algorithm is available via web-download, see 
	 Ref.~\cite{wwwaddress}.  
	 In order to demonstrate its reliability, we have simulated
	 the two-time correlation function of a driven 2-level atom. The coupling to the 
	 classical driving field is described by the Hamiltonian
   $
     \hat H \,=\, {1\over 2} \Omega \left(
        |e\rangle\langle g| + |g\rangle\langle e|
     \right)
   $.
   Spontaneous emission is described by the Lindblad operator 
   (\ref{simpleMasterEquation}), with $\hat\sigma=\sqrt{\gamma}|g\rangle\langle e|$.
	 In Fig.~\ref{mollowgt} we compare the results of our simulation with the exact
	 solution.
   We find excellent agreement between the results of the simulation and the exact
   formula for both the correlation function. A somewhat larger model system is
   considered in the next section.

\section{The optical parametric oscillator}
  \label{sectionExampleDOPO}

  We consider a system of two resonant optical modes $\hat a_1$ (fundamental)
  and $\hat a_2$ (second harmonic), interacting in a $\chi^{(2)}$-medium via the Hamiltonian
  \begin{equation}
	  \label{DOPOnonlinearH}
    \hat H_I \,=\, i {\kappa\over 2} \left(
      \hat a_1^{\dagger^2} \hat a_2 - \hat a_1^2 \hat a_2^\dagger
    \right)
    \,,  
  \end{equation}
  where $\kappa$ is the coupling constant.
  The second harmonic mode is pumped by a coherent field of amplitude $\epsilon$:
  \begin{equation}
	  \label{DOPOpumpH}
    \hat H_P \,=\, i \left( \epsilon \hat a_2^\dagger - \epsilon^\ast \hat a_2 \right)
    \,,
  \end{equation}
  and both modes are damped, with rates $\gamma_1$ and $\gamma_2$, respectively:
  \begin{equation}
	  \label{DOPOdampingL}
    {\cal{L}} \hat \rho \,=\,
      \gamma_1 \left(
        2 \hat a_1 \hat\rho \hat a_1^\dagger
          - \hat a_1^\dagger \hat a_1 \hat\rho - \hat\rho \hat a_1^\dagger \hat a_1
      \right)
      + \gamma_2 \left(
        2 \hat a_2 \hat\rho \hat a_2^\dagger
          - \hat a_2^\dagger \hat a_2 \hat\rho - \hat\rho \hat a_2^\dagger \hat a_2
      \right)
    \,.
  \end{equation}
  A detailed discussion of this system can be found in Ref.~\cite{milburn}.
  For pump field amplitudes above the threshold 
  $\epsilon_{\rm th} = \gamma_1\gamma_2 / \kappa$, there exist two classical steady
  state solutions for the fundamental mode amplitude:
  \begin{equation}
    \alpha_1 = \pm \sqrt{ {2\over\kappa} (\epsilon - \epsilon_{\rm th}) }
    \,.
  \end{equation}
  The quantum state resembles this classical solution: above threshold, the Wigner function
  consists of two peaks close to the classical solutions.
  The superposition of these peaks is incoherent, so the steady state is a classical
	mixture of two states both of which are well localized and have a well-defined quantum
	phase.
  However, tunneling between the two stable states is possible, corresponding to a 
  change of the phase angle by $\pi$. Therefore, above threshold the
  tunneling events will be the main reason for the decay of the correlation function
  $g(t) := { \left\langle \hat a_1^\dagger(t) \hat a_1 \right\rangle \over 
                  \left\langle \hat a_1^\dagger \hat a_1 \right\rangle } $
  in the long-time regime.
	Following \cite{drummond}, the tunneling is modeled to be a telegraph noise process with 
	a characteristic time $T$, which leads to an exponential decay
  \begin{equation}
    g(t) \,=\, e^{-{2\over T}t}
    \,.
  \end{equation}
	By adiabatic elimination of the second harmonic mode (the pump mode), which is a good
	approximation only in the case of a rapidly decaying pump mode, i.~e. for 
	$\gamma_2 \gg \gamma_1$, Kinsler and Drummond \cite{drummond} found the 
	following analytical expression for the tunneling time $T$:
  \begin{equation} \label{dkApprox}
    T \,\approx\, {\pi\over\gamma_1}
      \sqrt{ \lambda + \sigma \over \lambda (\lambda-\sigma)^2 }
      \exp\left\lbrace
        {2\over G^2} \left[
          \lambda - \sigma - \sigma \ln\left( \lambda\over\sigma \right)
        \right]
      \right\rbrace
    \,,
  \end{equation}
  where $\lambda := \epsilon / \epsilon_{\rm th}$ is the normalized pump field amplitude,
  $G := \kappa \sqrt{2\gamma_1\gamma_2}^{-1}$ is the scaled coupling constant, and
	$\sigma := 1 - G^2 / 2$.
  Eq.~(\ref{dkApprox}) is obtained in the ``potential-barrier approximation'' 
  which is valid in
	the limit of large threshold photon numbers, i.~e., for $G << 1$, and a large 
	potential barrier, i.~e., it fails for $\lambda$ very close to or below the 
	threshold $\lambda=1$.
	
	We carried out numerical simulations of the full quantum dynamics specified in
  Eqs.~(\ref{DOPOnonlinearH}-\ref{DOPOdampingL}), i.~e. without 
	adiabatic elimination of the pump mode, 
	using a truncated number state representation of the state vectors. 
 	Classical analysis predicts a photon number $N_2 = 1$  for the pump mode
	above threshold, and $N_1 = 8$ for the fundamental mode at $\lambda = 2$
	for the choosen values $\gamma_1 = \kappa,\,\gamma_2 = 4\kappa$.
	Since photon number fluctuations are enhanced by the nonlinear interaction, 
	the truncation values of a number state representation
	have to be chosen considerably higher than these values.
	We used truncation values for the fundamental mode 
	between a photon number $N_1 = 24$ for $\lambda \approx 1$, and $N_1 = 48$ for 
	$\lambda \approx 2$. 
	Tests with truncation values up to $N_1 = 64$ indicated that no significant error 
	was caused by this truncation.
	The Hilbert space of the second harmonic mode was truncated at a photon 
	number $N_2 = 16$.
  
\begin{figure}
    \epsfxsize=120mm\epsffile{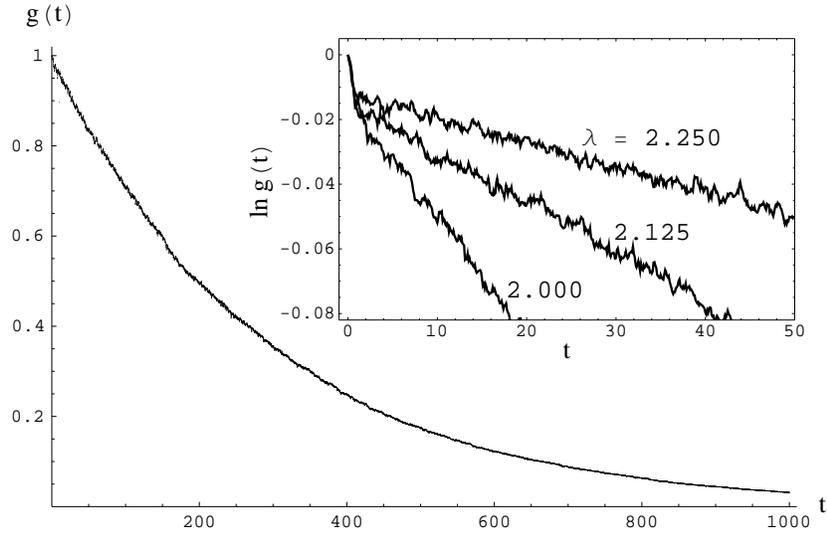}
    \caption{
		  The correlation function $g(t)$ of the optical parametric 
			oscillator for system parameters $\gamma_1 = \kappa = 1$, $\gamma_2 = 4\kappa$, 
			and a normalized pump amplitude 
			$\lambda = {\epsilon \over \epsilon^{\rm th}} = 2.0$, derived from 500 
			simulated trajectories.
			The inset displays a close-up view
			for different values of $\lambda$, showing the
      fast initial decay before the slow tunneling regime.
    }
    \label{dopocombi}
  \end{figure}

\begin{figure}  
    \epsfxsize=120mm\epsffile{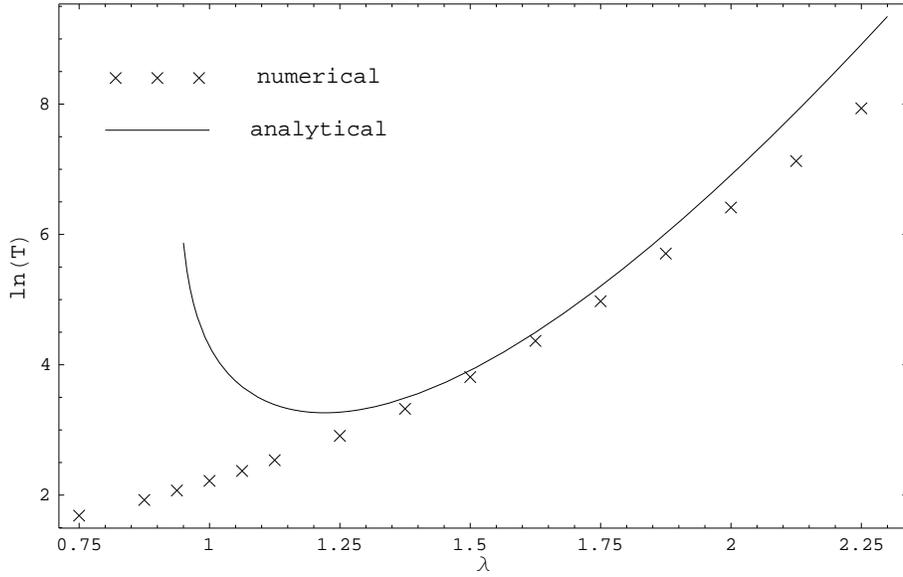}
    \caption{
      Comparison of numerical values for the tunneling times of the optical 
      parametric oscillator and the analytical approximation \ref{dkApprox},
      for system parameters as in Fig.~\ref{dopocombi}.
    }
    \label{dopologplot}
  \end{figure}

	Results of our numerical simulations of $g(t)$ are depicted in Fig.~\ref{dopocombi}
	and Fig.~\ref{dopologplot}. Quite generally, $g(t)$ displays a fast transient, which decays
	on a small time 
  scale~$\propto \kappa\epsilon \gamma_2^{-1} $%
	, followed by a slow exponential decay 
  $\propto \exp( -2t/T  )$ which governs the bevaviour of 
	$g(t)$ in the limit $t\longrightarrow\infty$.
	
  In Fig.~\ref{dopologplot} we depict the tunneling times $T$ 
  which we extracted from out numerical data as a function of the normalized
	pump field amplitude $\lambda$. We see a good agreement between our simulation and the 
	prediction of Eq.~(\ref{dkApprox}), for intermediate values of $\lambda$. For larger values
	of $\lambda$, the tunneling times predicted by our simulations are consistently shorter
	than predicted by Eq.~(\ref{dkApprox}). This may be due to the full quantum dynamics
	of the pump mode which is taken into account in our simulations.

\section{Summary}

 We have derived an efficient method for the numerical simulation of quantum
 mechanical two-time correlation functions 
 which is based on stochastic wave function propagation.
 In comparison with other algorithms, our algorithm was demonstrated to be
 generally more efficient, i.~e. requiring less runs for a reliable prediction
 of $g(t)$.
 We have successfully applied our method for the simulation of $g(t)$ of a nonlinear optical
 parametric oscillator.

 The tests indicate that our algorithm is stable and efficient, even for
 ``large'' (i.\,e., having large Hilbert spaces) problems:
 despite the seemingly more complicate pseudolinear equations of motion
 (\ref{geometricEquationOfMotion}),
 a proper implementation need not be slower but can be of
 of faster convergence than previously published algorithms.
 This stems from the fact that the most time consuming operations are the application
 of operators to vectors (operations of the type ``matrix times vector''), and all
 algorithms require the computation of 
 $\hat\sigma_k^\dagger \hat\sigma_k \left| \Phi \right\rangle$ and
 $\hat\sigma_k^\dagger \hat\sigma_k \left| \Psi \right\rangle$
 for every computation of a time derivative to be used in 
 the numerical integration of the continuous (no-jump) part of the evolution.
 All other quantities required in our algorithm can be obtained by computing scalar 
 products of vectors and purely scalar operations, which are comparatively cheap 
 operations.

\section{Acknowledgements}

 We gratefully acknowledge financial support for this work by the 
 Deutsche Forschungsgemeinschaft.
 We would like to thank Almut Beige, Patrick Navez, Jens Eisert and Carsten Henkel
 for helpful discussions.

\appendix
\section{The doubled-Hilbertspace method}
  \label{appendixBKP}

  The algorithm proposed in \cite{breuer2} is based on the observation, that 
  the problem of propagating the skew-symmetric $\hat\chi(t)$, as defined
  in Sec.~\ref{subsectionSkewProblem}, can be reduced
  to solving a master equation for a positive symmetric operator in a
  Hilbert space of twice the dimension of the original space. 

  The new density operator $\hat\Omega(t)$ in this doubled space is defined by 
  the initial condition
  \begin{eqnarray} \label{defOmega}
     \hat\Omega(0) &=&  \left( \matrix{ 
           \left| \phi(0) \right\rangle \left\langle \phi(0) \right|
           & \left| \phi(0) \right\rangle \left\langle \psi(0) \right|
       \cr
           \left| \psi(0) \right\rangle \left\langle \phi(0) \right|
           & \left| \psi(0) \right\rangle \left\langle \psi(0) \right|
     } \right)
  \end{eqnarray}
  and the equation of motion
  \begin{eqnarray}
    \label{doubleMaster}
     {d\over dt} \hat\Omega(t) \,=\, {\cal L} \,\hat\Omega(t)
       \,:=\, \left( \matrix{ L & 0 \cr 0 & L } \right) \, \hat\Omega(t)
     \,.
  \end{eqnarray}
  The superoperator ${\cal L}$ is again a Lindblad operator:
  \begin{eqnarray}
     {\cal L} \,\hat\Omega \,=\,
        2 \hat\Sigma \hat\Omega \hat\Sigma^\dagger 
           - \hat\Sigma^\dagger \hat\Sigma \hat\Omega
                  - \hat\Omega \hat\Sigma^\dagger \hat\Sigma
     \,,
  \end{eqnarray}
  where
  \begin{equation}
    \label{defSigma}
    \hat\Sigma \,:=\, \left( \matrix{ \hat \sigma & 0 \cr 0 & \hat \sigma } \right)
     \,.
  \end{equation}
  The initial value $\hat\Omega(0)$ can be written as a symmetric dyadic product,
  \begin{eqnarray}
     \hat\Omega(0) &=& \left| \Upsilon(0) \right\rangle \left\langle \Upsilon(0) \right|
     \:,
  \end{eqnarray}
  where 
  \begin{eqnarray}
     \left| \Upsilon(0) \right\rangle &=&  \left( \matrix{
           \left| \phi(0) \right\rangle
       \cr \left| \psi(0) \right\rangle
     } \right)
  \end{eqnarray}
  is a normalized pure state in the doubled Hilbert space. 
  Standard algorithms can now be used to unravel (\ref{doubleMaster}) into a 
  stochastic process for a state vector $\left| \Upsilon_r(t) \right\rangle$ 
  in this doubled Hilbert space, e.~g. the jump algorithm (\ref{simpleSymmetricJump})
  can be used:
  \begin{eqnarray}
    \label{kapplerJump}
       \left| d\Upsilon_r(t) \right\rangle
     \,=\, \left\lbrace \matrix{
         \left(
           - \hbox{\rm 1\hskip-0.3em l} + \left| \hat\Sigma \left| \Upsilon_r(t) \right\rangle \right|^{-1} \hat \Sigma 
         \right) \left| \Upsilon_r(t) \right\rangle
          ,&
             dp = 2 dt \left\langle \Upsilon_r(t) | \hat \Sigma^\dagger \hat \Sigma | \Upsilon_r(t) \right\rangle
       \cr
         \left(
           \left\langle \Upsilon_r(t) | \Sigma^\dagger\Sigma | \Upsilon_r(t) \right\rangle
           - \Sigma^\dagger\Sigma 
         \right) dt
           \left| \Upsilon_r(t) \right\rangle
         ,& 1 - dp
          \,.
       }
     \right.
  \end{eqnarray}
  The jump probability is computed as the arithmetic mean,
  \begin{eqnarray}
       dp &=& 2 dt \left\langle \Upsilon_r(t) | \hat \Sigma^\dagger \hat \Sigma | \Upsilon_r(t) \right\rangle
   \cr   &=& 2 dt \left(
       \left\langle \phi_r(t) | \hat \sigma^\dagger \hat \sigma | \phi_r(t) \right\rangle
       + \left\langle \psi_r(t) | \hat \sigma^\dagger \hat \sigma | \psi_r(t) \right\rangle
     \right)
     \,,
  \end{eqnarray}
  as opposed to the geometric mean suggested in our algorithm (\ref{ourJumpProbability}).
  However, unlike (\ref{GZparameters}), this algorithm is symmetric 
  with respect to the two components $ \left| \phi_r(t) \right\rangle$ and 
  $\left| \psi_r(t) \right\rangle$. 
  Moreover, the condition
  \begin{eqnarray} \label{doubleHilbertNorm}
     \left\langle \phi_r(t) \vert \phi_r(t) \right\rangle
     + \left\langle \psi_r(t) \vert \psi_r(t) \right\rangle
     \,=\, 1
     \,,
  \end{eqnarray}
  which is the proper normalization of vectors in the doubled Hilbert space, 
  will be fullfilled for all realizations, so exponential growth of the norm 
  of one of the vectors as in (\ref{gardinerExponential}) cannot 
  occur\footnote{
    Consequently, a constant factor must usually be applied when computing 
    expectation values, to compensate for this normalization.
  }. 

  Noteworthy, the condition (\ref{doubleHilbertNorm})
  is actually more restrictive than necessary:
  in the doubled Hilbert space, we are only interested in expectation values 
  of operators of the Form
  \begin{eqnarray} \label{lowerLeftOperator}
    {\cal A} \,=\, \left( \matrix{  0 & 0 \cr A & 0 } \right)
    \:;
  \end{eqnarray}
  in particular, the expectation value of the identity operator
  \begin{eqnarray}
    {\cal I} \,=\, \left( \matrix{ 
          \hbox{\rm 1\hskip-0.25em l} & 0 
      \cr 0 & \hbox{\rm 1\hskip-0.25em l} 
    } \right)
  \end{eqnarray}
  is irrelevant, and norm conservation in the stochastic average,
  \begin{eqnarray}
    \overline{
      \left\langle \phi_r(t) \vert \phi_r(t) \right\rangle
      + \left\langle \psi_r(t) \vert \psi_r(t) \right\rangle
    }
    \,=\, 1
    \,,
  \end{eqnarray}
  is not required. Therefore, condition~(\ref{doubleHilbertNorm}) is less 
  well justified than the corresponding 
  condition~(\ref{strictNormConservation}) in the unraveling of master equation 
  (\ref{simpleMasterEquation}) for a symmetric operator, and we believe 
  that it should be replaced by condition (\ref{ourCondition})%
  .

\section{The M\o{}lmer-Castin-Dalibard algorithm}
  \label{appendixMCD}

  Already in \cite{molmer1}, the following stochastic jump process
  was proposed (although formulated as in Eq.~(\ref{mcdOrig})):
  \begin{eqnarray} \label{mcd}
    \left( \matrix{ 
          \left| d\phi_r(t) \right\rangle 
      \cr \left| d\psi_r(t) \right\rangle 
    } \right) \,=\, \left\lbrace
      \matrix{
         \noalign{ \hrule width 0pt height 6pt }
      \cr
          \left(
            - \hbox{\rm 1\hskip-0.25em l} 
              + \left| \hat\sigma 
                  \left| \psi_r(t) + \nu \phi_r(t) \right\rangle
                \right|^{-1} \hat\Sigma
          \right) 
              \left( \matrix{ 
                    \left| \phi_r(t) \right\rangle 
                \cr \left| \psi_r(t) \right\rangle 
              } \right) 
           ,& \hskip-2mm
              dp \,=\, 2 dt \left| \hat\sigma 
                  \left| \psi_r(t) + \nu \phi_r(t) \right\rangle
                \right|^2
             \hskip-5mm
      \cr
         \noalign{ \hrule width 0pt height 4pt }
      \cr
        \left(
          \left| \hat\sigma | \psi_r(t) + \nu \phi_r(t) \rangle \right|^2
            - \hat\Sigma^\dagger\hat\Sigma 
        \right) dt
            \left( \matrix{ 
                  \left| \phi_r(t) \right\rangle 
              \cr \left| \psi_r(t) \right\rangle 
            } \right)
        , \hskip-3mm &\,  1 - dp
      \cr
         \noalign{ \hrule width 0pt height 8pt }
      }
    \right.
  \end{eqnarray}
  Here, $\nu$ is a phase factor with $|\nu| = 1$,
  $\hat\Sigma$ is the operator defined in Eq.~(\ref{defSigma}),
  and we use the abbreviation
  $\left| \phi_r(t) + \nu \psi_r(t) \right\rangle \,\equiv\,
    \left| \phi_r(t) \right\rangle + \nu \left| \psi_r(t) \right\rangle $.
  The vectors are normalized such that 
  $\left| \left| \phi_r(t) + \nu \psi_r(t) \right\rangle \right| \,\equiv\, 1$;
  it is easy to verify that the process (\ref{mcd}) will conserve this
  normalization. 
  In the original formulation of this method \cite{molmer1}, it was
  suggested not to propagate the pair 
  $\left( \left| \psi_r(t) \right\rangle, \left| \phi_r(t) \right\rangle \right)$, 
  but the
  linear combination
  $\left| \psi_{\nu,r}(t) \right\rangle \,:=\,
     \left| \phi_r(t) + \nu \psi_r(t) \right\rangle$; 
  this is possible, as all coefficients appearing in both branches of
  Eq.~(\ref{mcd}), as well as the jump probability $dp$, are functionals of
  $\left| \psi_{\nu,r}(t) \right\rangle $, and we can write:
  \begin{eqnarray} \label{mcdOrig}
      \left| d\psi_{\nu,r}(t) \right\rangle 
    \,=\, \left\lbrace
      \matrix{
        \left(
          - \hbox{\rm 1\hskip-0.25em l} 
          + \left| \hat\sigma \left| \psi_{\nu,r}(t) \right\rangle \right|^{-1} \hat\sigma
        \right) 
          \left| \phi_{\nu,r}(t) \right\rangle 
         ,&
            dp \,=\, 2 dt \left| \hat\sigma 
                \left| \psi_{\nu,r}(t) \right\rangle
              \right|^2
      \cr
        \left(
          \left\langle \psi_{\nu,r}(t) 
                | \hat\sigma^\dagger\hat\sigma | 
               \psi_{\nu,r}(t) 
          \right\rangle
            - \hat\sigma^\dagger\hat\sigma 
        \right) dt
          \left| \phi_{\nu,r}(t) \right\rangle 
        , \hskip-3mm &\:  1 - dp
      }
    \right.
  \end{eqnarray}

  However, merging of both vectors 
  $\left| \psi_r \right\rangle$ and $\left| \phi_r \right\rangle$ 
  into one only allows for a simulation of
  \begin{eqnarray}
    \hat\chi_\nu(t) &:=& \overline{
      \left| \psi_{\nu,r}(t) \right\rangle
      \left\langle \psi_{\nu,r}(t) \right|
    }
  \cr
    &=&
      \overline{
        \left| \psi_r(t) \right\rangle \left\langle \psi_r(t) \right|
      } + \nu \overline{
        \left| \phi_r(t) \right\rangle \left\langle \psi_r(t) \right|
      } + \nu^\ast \overline{
        \left| \psi_r(t) \right\rangle \left\langle \phi_r(t) \right|
      } + \overline{
        \left| \phi_r(t) \right\rangle \left\langle \phi_r(t) \right|
      }
    \,.
  \end{eqnarray}
  To extract the quantity of interest,
  $ \hat\chi(t) \,=\, \overline{ 
      \left| \phi_r(t) \right\rangle \left\langle \psi_r(t) \right|
    }
  $, 
  one has to perform the simulation for four different values of $\nu$:
  \begin{equation}
    \hat\chi(t) \,=\, {1\over 4} \left( 
      \hat\chi_{+1}(t) - \hat\chi_{-1}(t) 
         - i \hat\chi_{+i}(t) + i \hat\chi_{-i}(t)
    \right)
    \,.
  \end{equation}
  Since (\ref{mcdOrig}) is just the simple jump process 
  (\ref{simpleSymmetricJump}), applied to $|\psi_{\nu,r}(t)\rangle$,
  the problem of propagating the skew object $\hat\chi(t)$ has been reduced
  to the problem of unraveling 4 proper, positive definite density operators.

  Clearly, the extra work can be avoided (at the cost of having to propagate
  two separate vectors instead of a single one) by not merging the vectors and
  using the stochastic process (\ref{mcd}) directly.


\section{A concievable improvement which turns out to fail}
  \label{appendixFailure}

  The algorithm derived in Sec.~\ref{ourDerivation} was based on the requirement
  to simulate all operators of the type (\ref{lowerLeftOperator}), disregarding
  more general operators in the doubled Hilbert space.

  If all we want to compute is the expectation value 
  \begin{equation} \label{gtStoch}
    g(t) \,=\, \overline{
      \left\langle \psi_r(t) | \hat A | \phi_r(t) \right\rangle
    }
  \end{equation}
  for just {\sl one special} operator $\hat A$, we could ask whether one
  can formulate an even more specialized algorithm fitted to exactly this problem.
  More specifically, instead of the error (\ref{desiredCondition0}) of the estimate
  of the skew operator $\tilde\chi$, we could wish
  that the mean square of the error of our estimate of $g(t)$ should be 
  minimized. Again, we cannot obtain the global minimum and have to resort to a local
  condition, analogous to Eq.~(\ref{ourCondition}):
   \begin{equation} \label{specializedCondition}
     \forall \, \left| \phi \right\rangle, \left| \psi \right\rangle: \quad
     \overline{
       {d\over dt} \left|
         \left\langle \psi | \hat A | \phi \right\rangle
       \right|^2
     } \, \mathop=\limits^{!} \: {\rm minimal}
     \,.
   \end{equation}
   An algorithm fullfilling (\ref{specializedCondition}) can be derived in much the 
   same way as from Eq.~(\ref{ourCondition}). In particular, the corresponding jump 
   probability is:
   \begin{equation} \label{specializedJumpP}
     dp \, = \,
       2 \, \sum_{k=1}^n \left|
       {   \left\langle \psi_r | \hat\sigma_k^\dagger\hat A\hat\sigma_k | \phi_r \right\rangle
         \over
           \left\langle \psi_r | \hat A | \phi_r \right\rangle  }
     \right| \, dt
     \,.
   \end{equation}
   Jumps will not change the value of $|g_r(t)|$ in this algorithm\footnote{
     Even the phase of $g_r(t)$ can be conserved if we relax the requirement of our parameters
     $\eta_{1,\rm J}$ and $\eta_{2,\rm J}$ to be real-valued.
   }.
   At first glance, this method looks promising: 
   when applied to the simulation of the correlation function
   $g(t) = \left\langle \hat\sigma^\dagger\hat\sigma \right\rangle$ of a
   2-level atom, then we have to choose $\hat A \equiv \hat\sigma^\dagger$ and
   (\ref{specializedJumpP}) will contain a product $\hat\sigma\hat\sigma \equiv 0$. 
   Therefore, no jumps will ever occur for {\sl any} states 
   $\left| \phi_r \right\rangle$, $\left| \psi_r \right\rangle$, 
   and the algorithm becomes deterministic even for the driven 2-level atom. 
   Unfortunately, the results will be incorrect.
   In general, the method derived from condition~(\ref{specializedCondition})
   turns out to be highly unstable at best, and produces completely wrong results in some
   cases. This happens because 
   it does not posses a property analogous to (\ref{monotonousDecay}); rather,
   $\left| g_r(t) \right|^2$ may very well grow large for certain trajectories during the
   deterministic continuous evolution.
   Moreover, the algorithm will completely avoid to jump into states for which $g_r(t)$ vanishes, 
   although such states can be important as $g_r(t)$ may very well take on finite values at
   a later time.

\end{document}